\documentclass[pre,twocolumn,balancelastpage,groupedaddress,aps]{revtex4-1}

\usepackage{tikz}
\usepackage{dsfont}
\usetikzlibrary{matrix,arrows,decorations.pathmorphing}
\usepackage{amsfonts}
\usepackage{amsmath}
\usepackage{url}
\usepackage{graphicx}
\usepackage{amsthm}
\usepackage{amsmath}
\usepackage{color}
\usepackage{amssymb}
\usepackage{bm}
\usepackage{float}
\usepackage[unicode=true,pdfusetitle,
 bookmarks=true,bookmarksnumbered=false,bookmarksopen=false,
 breaklinks=false,pdfborder={0 0 1},backref=false,colorlinks=false]
 {hyperref}
 \def\normOrd#1{\mathop{:}\nolimits\!#1\!\mathop{:}\nolimits}

\begin{document}
\title{Resonances in a periodically driven bosonic system}
\author{Anton Quelle}
\affiliation{Institute for Theoretical Physics, Center for Extreme Matter and Emergent Phenomena, Utrecht University, Leuvenlaan 4, 3584 CE Utrecht, The Netherlands}
\author{Cristiane Morais Smith}
\affiliation{Institute for Theoretical Physics, Center for Extreme Matter and Emergent Phenomena, Utrecht University, Leuvenlaan 4, 3584 CE Utrecht, The Netherlands}
\begin{abstract}
Periodically driven systems are a common topic in modern physics. In optical lattices specifically, driving is at the origin of many interesting phenomena. However, energy is not conserved in driven systems, and under periodic driving, heating of a system is a real concern. In an effort to better understand this phenomenon, the heating of single-band systems has been studied, with a focus on disorder- and interaction-induced effects, such as many-body localisation. Nevertheless, driven systems occur in a much wider context than this, leaving room for further research. Here, we fill this gap by studying a non-interacting model, characterised by discrete, periodically spaced energy levels that are unbounded from above. We couple these energy levels resonantly through a periodic drive, and discuss the heating dynamics of this system as a function of the driving protocol. In this way, we show that a combination of stimulated emission and absorption causes the presence of resonant stable states. This will serve to elucidate the conditions under which resonant driving causes heating in quantum systems.
\end{abstract}
\maketitle
\section{Introduction}
Periodically driven systems are ubiquitous in physics. In optical lattices, the introduction of periodic driving allows one to modify the hopping parameters in the Hamiltonian in a variety of ways. Firstly, one can shake the lattice, which simply renormalises the hopping parameters \cite{Lignier2007,Tarruell2012}. This can be used to tune the behaviour of the Dirac cones in honeycomb lattices, for example \cite{Tarruell2012,Koghee2012}, but it can also be used to tune a superfluid-Mott insulator transition \cite{Eckardt2005,Creffield2006,Zenesini2009}. Shaking in a non-linear manner does not just change the hopping amplitude, but also impinges a phase to the hopping term, which corresponds to the generation of an effective gauge field to the lattice. This technique was first used to simulate classical frustrated magnetism in a triangular optical lattice filled with bosons \cite{Struck2011}, and later it has been, together with laser-assisted tunneling through the use of Raman excitations, a very efficient tool to realise fermonic topological phases with ultracold atoms in optical lattices \cite{Beaufils2011}. 

Recent studies have shown that it is possible to generate even more complex scenarios in driven systems. An interesting possibility is the shaking of Feshbach resonances, which induces a renormalisation of the hopping parameter in a way that depends on the density operator in the initial and final sites. The original theoretical predictions by Greschner et al. \cite{Greschner2014} for bosons have been recently experimentally verified \cite{Meinert2016}. For fermions, this procedure allows for the quantum simulation of a well-known model studied in the context of high-Tc superconductors, the so-called correlated hopping model \cite{Liberto2014}. 

Despite these efforts, shaking remains the most common tool in optical lattices \cite{Eckardt2016}. One can distinguish three separate shaking regimes \cite{Quelle2016}. Firstly the so-called quasi-equilibrium regime, where $J\ll\hbar\omega\ll\Delta$; here $J$ is the hopping parameter, which is roughly the bandwidth of the relevant set of bands, $\omega$ the shaking frequency, and $\Delta$ the gap to the nearest set of bands. It is the regime that has been most studied \cite{Eckardt2005,Arimondo2009,Koghee2012,Sengstock2013,Goldman2014}. Secondly, the regime where $J\ll\hbar\omega\sim \Delta.$ This regime is starting to attract interest in optical lattices \cite{Cheng2013,Zhai2014,Zhai2014PRA}. Finally, there is the regime where $J\sim \hbar\omega\ll\Delta,$ in which resonances due to the driving play a major role in the dynamics of the relevant bands, and Floquet topological phases can be realised \cite{Lindner2011}.

However, driven optical lattices heat \cite{Gerbier2010,Bilitewski2015}, which puts restrictions on the possible duration of experiments. The reason why the first regime described above is so much better explored, is due to the absence of resonances, which lessens heating. This process of heating is a generic feature of driven systems, although the system can be stable for very long time scales in the many-body case \cite{Maricq1982,Bukov2015,Kuwahara2016,Machado2017}. Especially if the underlying system is ergodic, the driving is expected to heat the system to infinite temperature \cite{DAlessio2014,Ponte2015}. Nevertheless, not every system is ergodic, and a great deal of research has been done on the heating behaviour within a band structure in the presence of interactions, disorder, and many-body localisation \cite{DAlessio2013,Bukov2016,Khemani2016,Weidinger2016}.

It is known that the presence of symmetries will restrain the level statistics of the heated system \cite{Regnault2016}, which is taken to the extreme in the case of integrability. Without integrability, however, systems usually thermalise because the driving couples large numbers of different energy states \cite{Lazarides2014}. A notable exception is the phenomenon of freezing, which has been predicted theoretically \cite{Das2010,Mondal2012}, and observed experimentally \cite{Hegde2014}, in some systems with bounded spectrum. However, one would expect that systems with a spectrum unbounded from above tend to accept energy indefinitely. Indeed, bound states tend to be more unstable in driven systems with unbounded energy, even without interactions \cite{Mori2015}. Nevertheless, the driving might couple different modes in a controlled manner, and this can prevent the system from thermalising. A canonical example of this behaviour is the quantum kicked rotor. For certain driving protocols, the model can be mapped onto the Anderson model \cite{Grempel1984}, and the driving prevents a delocalisation of the system in energy space due to Anderson localisation.

In the above example, the localisation in energy space is a consequence of disorder in the kinetic energy. However, it is possible to create a system that also has discrete, periodic and unbounded spectrum, but which localises in energy space without resorting to randomness. In this paper, we introduce such a driven model, in which most of the Floquet modes are local in energy space. This behaviour is a consequence of choosing a driving protocol that features notable stimulated emission of energy. Since Floquet modes do not gain/lose energy over a cycle, this means that the system hosts a large number of states that do not heat. We show which states do and do not heat, and discuss how to separate them by a proper choice of initial condition. This ties into the discussion above, since the driving protocol is perfectly resonant, showing that it is, in principle, possible to drive a system at resonance without pumping heat into it.

The layout of this paper is as follows. In Sec.~\ref{pumping}, we discuss some generalities about Floquet systems, and how the expected energy in a system can change during a period of the driving. In Sec.~\ref{undriven}, we introduce a one parameter family of Hamiltonians, inspired by a massless boson in a finite one-dimensional wire with a boundary mass at the edges. We solve these Hamiltonians, and construct the unitary operator relating the Hilbert spaces at different boundary mass. Using these results, we construct a driven system in Sec.~\ref{driving}, by varying the boundary mass in time. We analyse the resulting Floquet system for two different choices of parameters to demonstrate the features of the model. In Sec.~\ref{stability}, we discuss how the results from the previous section depend on the choice of parameters, and what this means for the stability of the system. Finally, we provide a summary and outlook in Sec.~\ref{conclusion}.

\section{Floquet systems and energy pumping}\label{pumping}

Before we introduce the model to be treated in this paper, we summarise some important results of the Floquet framework used to analyse time-periodic out-of-equilibrium systems. Then, we introduce the energy-pumping operator and discuss some of its useful properties.

In a Floquet system, the Hamiltonian is explicitly time-dependent in a periodic manner, so energy is no longer conserved, and there are no longer any stationary states in the system. However, due to the periodicity of the Hamiltonian, the propagator is also periodic, $U(t+T,t'+T)=U(t,t')$ and therefore $U(nT,0)=U(T,0)^n.$ Floquet theory is the analysis of the propagator $U:=U(T,0)$, which gives full information about the stroboscopic behaviour of the system. Because energy is not conserved in a Floquet system, such systems can exhibit a richer behaviour than equilibrium systems.  This can already be seen in the stroboscopic behaviour of the system, and hence in the spectral behaviour of $U$.  The simplest case is when $U$ has a complete basis of normalisable eigenstates, and the Hamiltonian is bounded. Because $U$ has a basis of normalisable eigenstates, the system repeatedly returns arbitrarily close to its initial state; since the Hamiltonian is bounded, the energy does so as well, and the system is stable \cite{Hogg1982}.

If the Hamiltonian is unbounded, or the propagator has unnormalisable eigenstates (i.e. states of infinite norm), the energy in the system may increase indefinitely, and more complicated behaviour can occur. In order to quantify this analysis, it is useful to consider the energy pumping over a single cycle, which is given by the operator
\begin{align}
	\Delta E=U^{-1}HU-H.
\end{align}
If $v$ is an eigenvector of $U$, and both $v$ and $Hv$ are normalisable, then $\langle v|\Delta E |v\rangle=0$. This implies that if $U$ is diagonalisable with finite energy states (for example if $U$ is diagonalisable and the Hamiltonian is bounded), then $\operatorname{Tr}(\Delta E)=0$, which follows by evaluating the trace using the eigenstates of $U$. This argument fails if there are eigenstates $v$ of $U$ for which either $\langle v|v\rangle$ or $\langle v|H|v\rangle$ is infinite. In this case, $\langle v|\Delta E |v\rangle\neq0$ is possible, and there exist steady states that emit or absorb energy indefinitely. Note that this is possible because these states are at infinite energy. We conclude that a Floquet system can only be stable if $\operatorname{Tr}(\Delta E)=0$. 

We will show below that since $\Delta E$ contains detailed information about the energy pumping of the system, it can be a useful tool in analysing the stability of Floquet systems. Because $\Delta E$ is Hermitian (while $U$ is not), numerical methods adapted specifically to such matrices can be applied. 

For the system we consider below, $\operatorname{Tr}(\Delta E)\neq0$, so it must have states that are not stable in energy. We use the information provided by $\Delta E$ to identify the problematic states, and find conditions under which the system can be stably driven.

\section{The energy Eigenstates}\label{undriven}

In this paper, we consider a non-interacting boson field $\phi$ defined on a line segment of length $l$, with a boundary mass $\mu$ on the edges that we will modulate in time. At a fixed time, the field theory has the Lagrangian
\begin{align}\label{Lagrangian}
	L_\mu(t)=\frac{\hbar u}{2 l}\int_0^l& \left\{\frac{l}{u^2}\phi_t^2(\bm x)-l\phi^2_x(\bm x)\right.\\ &\left.+\mu \phi^2(\bm x)\left[\delta(x-l)-\delta(x)\right]\right\}dx \nonumber
\end{align}
where subscripts denote derivatives, and $u$ is a characteristic velocity. Furthermore, $\bm x=(x,t)$ to make the notation more concise. Interacting versions of this Lagrangian have been used to model a variety of boundary effects, such as boundary critical behaviour \cite{Diehl1997} and the thermal Casimir effect \cite{Diehl2011}. One can also consider this Lagrangian to be a continuum approximation of the Bose-Hubbard model, as discussed in Refs~\cite{Eckardt2005,Creffield2006,Zenesini2009,Struck2011} in the limit of linear dispersion. The boundary mass term can then be considered to describe the effect of coupling the system to a reservoir. The convention is such that the field $\phi$ and the boundary mass $\mu$ are dimensionless. To properly define the action, we need to define the boundary conditions for the field and implement them. The most sensible boundary condition for the field is the Neumann boundary condition $\phi_x(0,t)=\phi_x(l,t)=0$, for reasons that will become clear below.

In order to quantise the theory, we first determine the structure of the classical system. The variation of the action reads
\begin{align*}
	\delta S_\mu=&\frac{\hbar u}{l}\int\delta \phi(\bm x)\left\{-\frac{l}{u^2}\phi_{tt}(\bm x)+l\phi_{xx}(\bm x)\right.\\ &\left.+\mu \phi(\bm x)\left[\delta(x-l)-\delta(x)\right]\right\}d\bm x +...
\end{align*}
, where the ellipsis denote a boundary term, coming from an integration by parts, that does not contribute to the equations of motion. From the bulk term, we read off
\begin{align*}
	\hbar^2\left\{\phi_{tt}(\bm x)-u^2\phi_{xx}(\bm x)-\frac{u^2}{l}\mu \phi(\bm x)\left[\delta(x-l)-\delta(x)\right]\right\}=0.
\end{align*}
Note that we have multiplied both sides of this equation with a non-zero prefactor. To find the independent dynamical degrees of freedom, we decompose the spatial part of this into eigenfunctions
\begin{align}\label{EigenEquation}
	- f_n''(x)-\frac{\mu}{l} f_n(x)\left[\delta(x-l)-\delta(x)\right]=\frac{E^2_n}{\hbar^2 u^2} f_n(x),
\end{align}
where the primes denote derivatives with respect to $x$, and $E_n$ is the energy of the system. We obtain from the equations of motion
\begin{align*}
	E_n^2\int_0^\epsilon f_n(x)dx&=\hbar^2 u^2\int_0^\epsilon\left(\frac{1}{l} \mu f_n(x)\delta(x)- f_n''(x)\right)dx\\
	&=\frac{\hbar^2 u^2}{l}\mu f_n(0)-\hbar^2 u^2 f_n'(\epsilon).
\end{align*}
By sending $\epsilon\downarrow 0$ and using the Neumann boundary condition, we obtain $l f_n'(0)=\mu f_n(0)$. We can repeat this argument at $x=l$, so that Eq.~\eqref{EigenEquation} reduces to $-\hbar^2 u^2 f_n''(x)=E^2_n f_n$ with boundary condition $l f_n'(0)=\mu f_n(0)$ and $l f_n'(l)=\mu f_n(l)$. The effect of the boundary term is to change the boundary condition in the Lagrangian, with $\mu=\infty$ corresponding to Dirichlet boundary conditions.
From this, we can obtain the functions $f_n$ solving Eq.~\eqref{EigenEquation}:
\begin{align*}
	f_0(x)&=\sqrt{\frac{2}{l} \frac{\mu}{\exp(2\mu)-1}}e^{\mu x/l}\\
	f_n(x)&= \frac{\sqrt{2}}{\sqrt{1+\left(\frac{\mu }{n \pi}\right)^2}}\left[\cos\left(\frac{\pi}{l}n x\right)+\frac{\mu}{n \pi}\sin\left(\frac{\pi}{l}n x\right)\right].
\end{align*}
These solutions are orthonormal with respect to $d\mu_l=dx/l.$ Since the Laplacian is Sturm-Liouville on the interval with Robin boundary conditions, these solutions form a complete basis. The corresponding energies are
\begin{align}
	E_0^2=-\left(\frac{\mu\hbar u}{l}\right)^2,\ \  E_n^2=\left(\frac{n\pi \hbar u}{l}\right)^2.
\end{align}
For $\mu\rightarrow 0$, the solutions just become cosines, corresponding to the Neumann boundary conditions, and $f_0$ is just the constant solution at zero energy. In general, $f_0$ is an exponentially decaying state bound to the edge due to the negative mass term there. For $\mu\rightarrow \infty$, $f_0$ becomes proportional to a delta function, which can be expressed in terms of the other solutions, which are sines corresponding to Dirichlet boundary conditions. Consequently, $f_0$ decouples from the system at $\mu=\infty$.
We can expand the field $\phi$ on these eigenfunctions $\phi(\bm x)=\sum_n \phi_n(t) f_n(x)$, to rewrite the Lagrangian as
\begin{align}\label{ExpLag}
L_\mu=\frac{1}{2}\sum_n \frac{\hbar l}{u}\dot \phi_{n}^2-\frac{E_n^2 l}{\hbar u} \phi_n^2.
\end{align}
Since we no longer have spatial derivatives appearing, we revert to an overdot for time derivates from now on. The Lagrangian in Eq.~\eqref{ExpLag} does not yield a model with a well defined particle number in the standard canonical quantisation. This is because the $n=0$ mode has imaginary frequency, yielding an inverted harmonic oscillator in the canonical quantisation procedure \cite{Barton1986}. Since the inverted harmonic oscillator does not have bound states, this situation is clearly unsatisfactory. On the other hand, the path integral quantisation for this system yields exponentially decaying correlation functions for $\phi_0$, due to the imaginary frequency. In the canonical picture, these exponentially decaying correlations correspond to so-called Gamow states \cite{Bohm1989,Schulte1995,Castagnino2000}, which are non-normalisable, exponentially decaying states that describe unstable modes in the system. In this picture, we can interpret the boundary mass in Eq.~\eqref{Lagrangian} as an effective term after integrating out a system coupled to the line segment. The state $\phi_0$ then describes a state bound to the edge of the line, which decays into the coupled system at an exponential rate.

We quantise the $n=0$ mode in this way, and all the other modes in the usual manner. The Hamiltonian then reads
\begin{align}\label{Hamiltonian}
	H_\mu&=\sum_n E_n a^\dagger_n a_n,
\end{align}
where 
\begin{align*}
	a_n=\frac{1}{\sqrt{2}}\left(\sqrt{\frac{E_n l}{\hbar u}}\phi_n +i \sqrt{\frac{u}{E_n \hbar l}} \pi_n\right).
\end{align*}
Here, $\pi_n$ is the canonical momentum conjugate to $\phi_n$, making $a_n$ an annihilation operator, and $E_0$ is purely imaginary. Because the system consists of non-interacting bosons, we now project onto the single-particle subspace $\mathcal{H}$, since this completely determines the dynamics of the many-particle system. It should be noted that $\mathcal{H}$ is equal to the span of the functions $f_i$, which are now interpreted as the wavefunctions of a particle, and that this span is independent of $\mu$.

We will give the overlap of the various basis elements of $\mathcal{H}$ corresponding to different $\mu.$ To indicate the value of the boundary mass, we now add superscripts $\mu,\nu$ to make the dependence of the basis on the boundary mass explicit; we will also introduce such a superscript in the Hamiltonian. The overlap between the different basis vectors reads:
\begin{align*}
\langle f_0^\mu|f_0^\nu\rangle&=\frac{e^{\mu+\nu}-1}{\mu+\nu}\sqrt{\mu(\coth\mu-1)}\sqrt{\nu(\coth\nu-1)},\\
\langle f_m^\mu|f_0^\nu\rangle&=\frac{\sqrt{2}\left((-1)^me^\nu-1\right)\pi(\nu-\mu)\sqrt{\nu(\coth\nu-1)}}{\sqrt{\pi^2+\mu^2/m^2}(\pi^2 m^2+\nu^2)},\\
\langle f_m^\mu|f_m^\nu\rangle&=\frac{\pi^2 m^2+\mu\nu}{\sqrt{\pi^2m^2+\mu^2}\sqrt{\pi^2m^2+\nu^2}},\\
\langle f_m^\mu|f_n^\nu\rangle&=\frac{2 \left(1-(-1)^{m+n}\right)(\mu-\nu)}{(m^2-n^2)\sqrt{\pi^2+\mu^2/m^2}\sqrt{\pi^2+\nu^2/n^2}}.
\end{align*}
From these expressions, we can show that if either $m$ or $n$ is large, $\langle f_m^\mu|f_n^\nu\rangle\rightarrow \delta_{m,n}$, so the modes decouple at high frequencies. For $\langle f_m^\mu|f_n^\nu\rangle$, this follows from the inequality $|m^2-n^2|>2\operatorname{min}(|m|,|n|)$. As we will see, this behaviour is crucial for the existence of stable states in the system.

\section{Effects of driving}\label{driving}

Now that we have analysed the Hamiltonian at a single instant in time, we can formally introduce our driving protocol. In terms of the Hamiltonians $H^\mu$, it reads:

\begin{align}\label{DrivingHam}
	H=\left\{\begin{aligned}
&H_\mu,& t\in[0,T_1)\  \operatorname{mod}(T)\\
&H_\nu,& t\in[T_1,T_1+T_2)\  \operatorname{mod}(T)
	\end{aligned}\right.
\end{align}
where $T=T_1+T_2$ is the driving period. Since the Hamiltonian, by construction, obeys $H(t+T)=H(t)$, we can use Floquet theory to get information from the system. Accordingly, we analyse the spectrum of the propagator
\begin{align}\label{Propagator}
	U=\sum_{m,n,o}e^{i(E_n T_1+E_o T_2)/\hbar}|f^\mu_m\rangle\langle f_m^\mu|
	f_o^\nu\rangle\langle f_o^\nu|f_n^\mu\rangle\langle f^\mu_n|.
	\end{align}
It should be noted that unless $\mu,\nu$ are either $0$ or $\infty$, $U$ contains an exponentially decaying mode, making the propagator non-unitary. This is consistent with the interpretation of $\phi_0$ as an unstable mode that decays out of the system, which causes probability to be non-conserved.

\subsection{Conformal limit}\label{conformal}
An interesting warmup case is the limit ${\mu\rightarrow\infty}$, ${\nu\rightarrow 0}$, so that $U$ is unitary, supplemented by the choice $T_1=T_2=T/2,$ so that we can write $E_n T_i=n \pi\theta \hbar$ for $\theta=uT/2l$. For these parameters the system is conformally invariant, since $\theta$ does not change under scaling of $T,l$, and the Floquet spectrum is expected to exhibit periodicities. In this case, the overlap of the basis elements becomes
\begin{align*}
\langle f_0^\infty|\psi\rangle&=0,\ \forall \psi\in\mathcal{H},\\
\langle f_m^\infty|f_0^0\rangle&=\frac{\sqrt{2}[1-(-1)^m]}{\pi m},\\
\langle f_m^\infty|f_n^0\rangle&=\left\{\begin{aligned}
\frac{2m[1-(-1)^{n+m}]}{\pi(m^2-n^2)},\ \ m\neq n,\\
0\ \ \ \ \ \ \ \ \ \ \ \ ,\ \ m=n .
\end{aligned}\right.
\end{align*}
Substituting these expressions into Eq.~\eqref{Propagator}, we find that the propagator splits up into an odd and an even part,
\begin{align*}
	U=\sum_{m,n}U^o_{m,n}|f^0_{2m-1}\rangle\langle f^0_{2n-1}|+\sum_{m,n}U^E_{m,n}|f^0_{2m}\rangle\langle f^0_{2n}|.
\end{align*}
As we demonstrate in Appendix~\ref{As}, $U^o$ and $U^e$ can be calculated in closed form due to the conformal symmetry. The asymptotic form as $n\rightarrow \infty$ is especially simple:
\begin{align}\label{ConProp}
	U^{o,e}_{n+a,n}=2\theta e^{i\pi \theta(4n+a)}\operatorname{sinc}(a\pi\theta),
\end{align}
which is valid for $a\neq 0$, and holds up to a global phase. The diagonal elements, with $a=0$, can be extracted by requiring unitarity. In Eq.~\eqref{ConProp}, a symmetry is present: a change in $n$ only modifies the phase, meaning that all the columns of the propagator, and hence the eigenvectors, are related by a shift operator in the asymptotic limit. This is due to the conformal symmetry, which causes all energy levels to be equivalent as soon as one forgets about the ground state.

From Eq.~\eqref{ConProp}, the asymptotic behaviour as $n\rightarrow\infty$ of $\langle f^0_n|\Delta E|f^0_n \rangle$ can be calculated, and it is infinite for all $n$, implying that the system is completely unstable as a consequence of the scaling symmetry. From a mathematical point of view, we see that all eigenstates of $U$ have infinite norm, which implies that $\operatorname{Tr}(\Delta E)=\infty.$ This is consistent with the discussion in Sec.~\ref{pumping}, where this was pointed out as one of the possible causes of instability. This is a result of choosing $\nu=\infty,$ which physically corresponds to an infinite driving strength.

\subsection{Finite driving strength}\label{finite}

Interestingly, by choosing $\nu$ in Eq.~\eqref{DrivingHam} to be finite, this problem is solved at the cost of manually breaking the scaling symmetry, and introducing the decaying mode. It turns out that in this case, there are many energetically stable modes that lose as much energy to stimulated emission as they gain through absorption. Physically, this can be traced back to the Lagrangian in Eq.~\eqref{Lagrangian}. The energy flux at space-time point $\bm x$ is given by the stress energy tensor $T_{t,x}(\bm x)=\normOrd{\pi(\bm x)\phi_x(\bm x)}$, where $\normOrd{...}$ denotes normal ordering, and the subscripts do not denote derivatives but tensor indices. As we show in Appendix~\ref{T}, the flux into the system is proportional to
\begin{align*}
\sum_{n,n'}\sqrt{\frac{ n'}{n}}\left[f_n(x) f_{n'}(x)\right]^{l}_{0}\left(a^\dagger_n a^\dagger_{n'} -a_n a_{n'}\right).
\end{align*}
This shows that the energy flux into the system vanishes in the single-particle sector. If this term were non-zero, there would be an energy flux in addition to that caused by switching the boundary mass, which would make the existence of stable states impossible.

\begin{figure}[hbt]
\includegraphics[width=\linewidth]{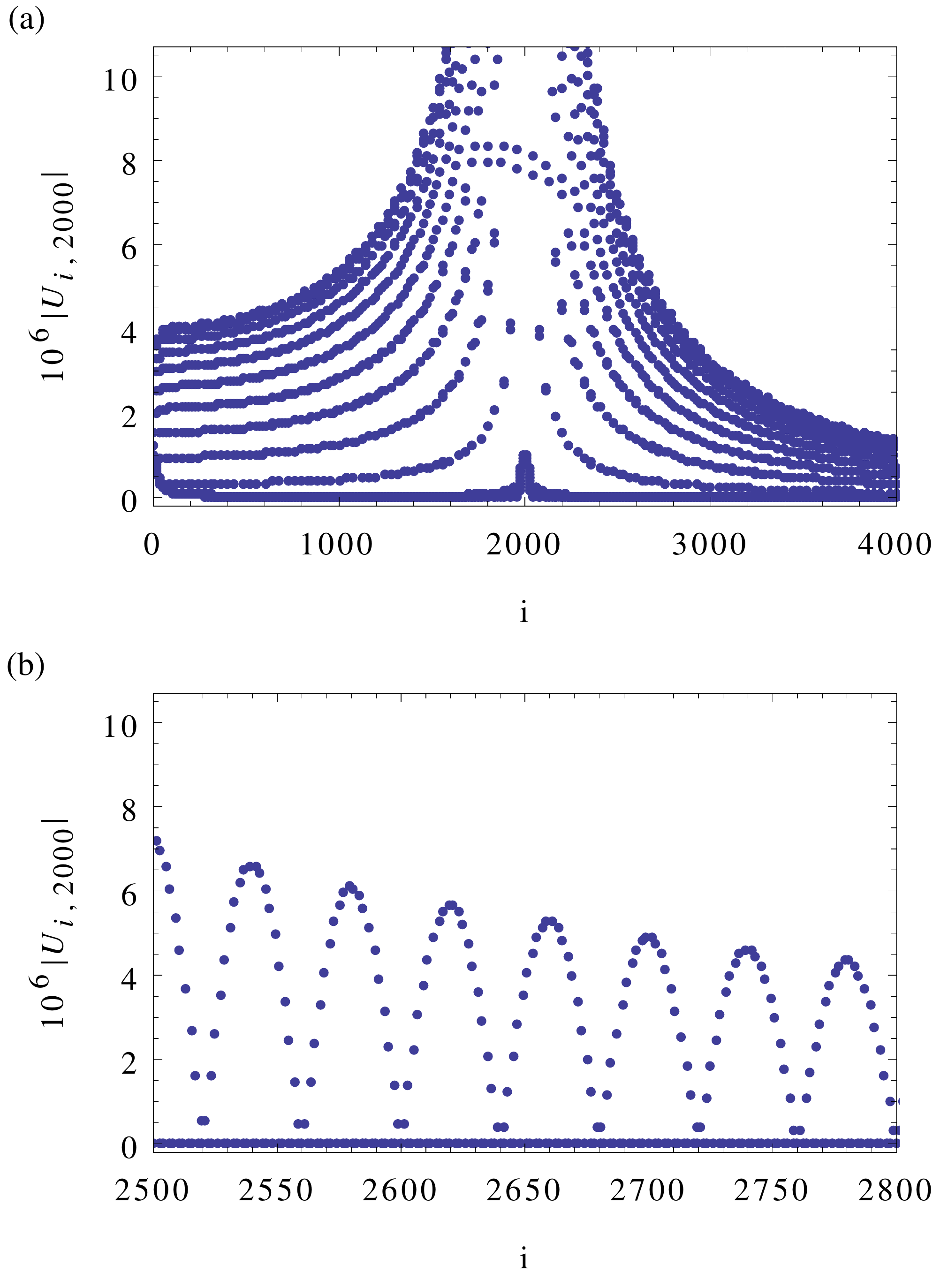}\caption{\label{PropCol2000} (a) The value $|U_{i,2000}|$ of the propagator (in units of $10^{-6}$) for $\mu=0$, $\nu=20,$ $\theta=1/20$, $M=1$. This is the absolute value of the $2000^{th}$ column of the propagator. The highest weight is at $i=2000$ since the propagator becomes diagonal at large frequencies. The overlap with the low energy states is higher than with the high energy states, which offsets the infinite tail on the right side. (b) Same as in (a) but for a restriced range of values so that the resonance structure is clearly visible. The peaks have a spacing of 40, which is precisely twice the driving frequency.}
\end{figure}

Because the sum over $o$ in Eq.~\eqref{Propagator} cannot be performed in closed form for finite $\nu$, we will rely on numerics. We choose the values $\mu=0,\nu=20$ for the two boundary masses. For the driving phase we use $\theta=1/20.$ Finally, we add a mass term
\begin{align*}
	-\frac{\hbar u}{2 l^2} M \int_0^l \phi^2(\bm x) dx
\end{align*}
to the Lagrangian. In this convention, $M$ is dimensionless, and we choose $M=1$. This does not change the eigenfunctions, but it sends $n\pi\mapsto \sqrt{n^2\pi^2+1}$ in the expression for $E_n$, which has the effect of removing the perfect periodicity in the energy spectrum for the lowest few energy modes (since $M=1$, the effect quickly becomes negligible for increasing $n$). Breaking the periodicity improves the convergence of the numerics, as we discuss in Sec.~\ref{stability}. Note that because $\mu>1$, the lowest energy mode still has imaginary energy, so the fundamental behaviour of the system is not altered. The absolute value of the $2000^{th}$ column of the propagator is shown in Fig.~\ref{PropCol2000} for these parameter values.

It can be seen that the highest weight is at value $U_{2000,2000}.$ This is consistent with the observation that $\langle f_m^\mu|f_n^\nu\rangle\approx \delta_{m,n},$ since it implies that the propagator becomes diagonal at high energies. The propagator has a higher weight on the low-energy states than on the high-energy states. Furthermore, a clear resonance structure is present in the propagator elements, which can be most clearly seen in Fig.~\ref{PropCol2000}(b). It should be noted that the peaks have a spacing of $40$, which is equal to $2/\theta$. Since $1/\theta=\hbar \omega/(E_n-E_{n-1})$, i.e. the driving frequency in units of the level spacing, these are driving resonances expected in any Floquet system. These resonances occur at twice the driving frequency because of an approximate parity symmetry. In the conformal case, this symmetry was exact, and resulted in the splitting of the Floquet propagator into an odd and an even part. In this case, the symmetry is slightly broken due to the finite boundary mass, but at large momenta the even and odd modes still decouple, as can be seen in Fig.~\ref{PropCol2000}.

The energy pumped into state $|f^0_{m}\rangle$ over a cycle is given by 
\begin{align}\label{EnergyPump}
\Delta E_m:=\langle f_m^0|\Delta E|f_m^0\rangle=\sum_i \left(E_i |U_{i,m}|^2\right)-E_m,
\end{align}
As we will discuss below, this sum is negative, except at low column numbers. Note that because we calculate this quantity for $\mu=0$, for which $E_0=0$, this quantity is real, and can be interpreted as the energy pumped over a cycle, as usual. For this to be true, the propagator has to have higher weight on low energy states, to offset the infinite tail at high energies, a property which we remarked.

\begin{figure}[hbt]
\includegraphics[width=\linewidth]{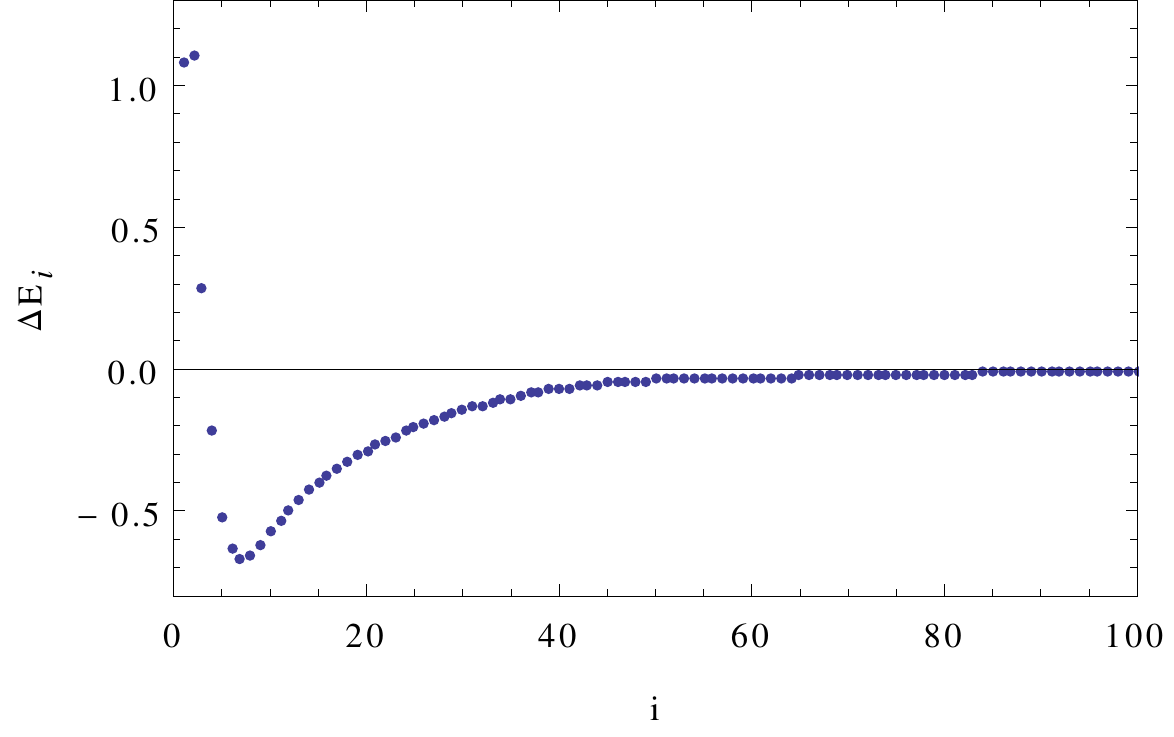}
\caption{\label{dE} The value of $\Delta E_i$ for the first $100$ values of $i$. It can be seen that $\Delta E_i$ is positive for $i<4$, and is negative for larger $i$. Although $\Delta E_i$ stays negative for large $i$, it asymptotically tends towards zero.}
\end{figure}

\begin{figure}[hbt]
\includegraphics[width=\linewidth]{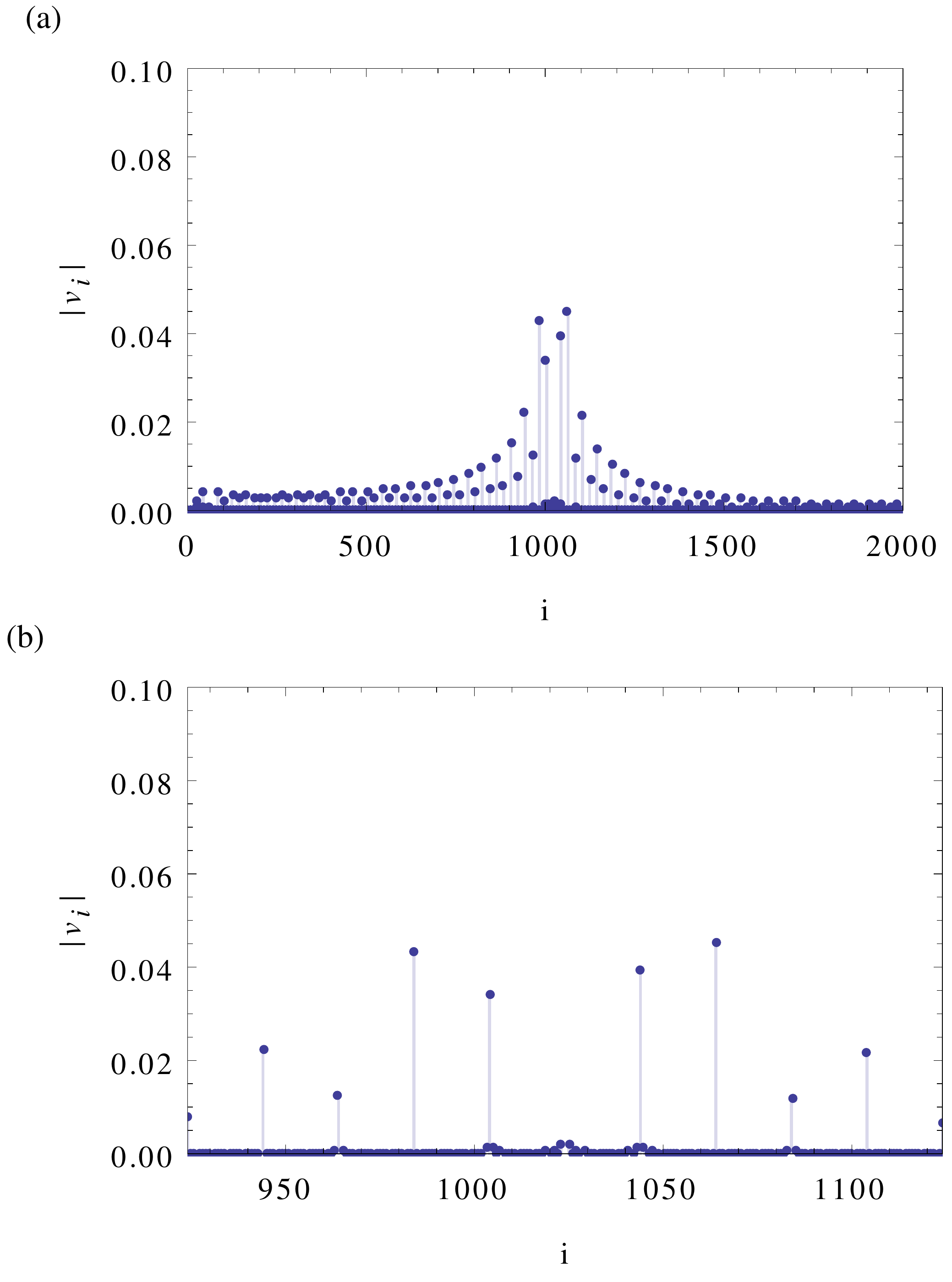}
\caption{\label{UNev} (a) The components $|v_i|$ of the $1024^{th}$ eigenvector $v$ of $U$ for $\mu=0$, $\nu=20,$ $\theta=1/20$, $M=1$. The corresponding eigenvalue has unit norm. The vertical lines serve as a guide to the eye, and show that the function has extremely sharp peaks, with most weights indistinguishable from zero. (b) A zoom of (a) to show the small scale structure. A smaller range of components is depicted, and the matrix elements are multiplied by $10$, so the scales are comparable. The decay of the peaks is now visible, but still rapid.}
\end{figure}

We have numerically calculated $U_{i,j}$ for $i,j\leq 4000,$ and from this we can approximate $\Delta E_m$ simply by truncating the sum at $4000$. The result is shown in Fig.~\ref{dE}. We see that the result is positive for small $m$, then becomes negative and tends to zero from below. Numerical results indicate that the summand $E_i |U_{i,m}|^2$ of $\Delta E_m$ decays as $1/[10(i-m)^2]$. Therefore, we can find an upper bound for $\Delta E_m$ by considering
\begin{align*}
 	\Delta E_m=\sum_i E_i |U_{i,m}|^2\leq \sum_{i=0}^N E_i |U_{i,m}|^2+\sum_{N+1}^\infty \frac{1}{10 (n-m)^2}.
 \end{align*} 
We have checked for several large $m$ that this expression becomes negative if $N$ is large enough. Therefore, the general behaviour in Fig.~\ref{dE} should be correct, although $\Delta E_m$ tends to zero slightly faster than shown in the figure; the numerical expressions go faster than $-1/m$. This implies that $\Delta E$ has a negative, but finite, trace. As we discussed in Sec.~\ref{pumping}, if $U$ is unitary, this can be because $U$ has eigenstates $v$ for which either $\langle v|v\rangle=\infty$ or $\langle v|H|v\rangle=\infty$. Here, the propagator $U$ is not unitary, due to the imaginary $E_0$, which in itself causes $\operatorname{Tr}(\Delta E)<0$ due to decay of the zero mode. However, in addition, there are eigenstates of $U$ for which $\langle v|H|v\rangle=\infty$, which causes further deviation from zero. Consequently, the decaying mode is not the only instability in the system, and there are states which will run off to large energy before they slowly decay out of the system. If one were to stabilise the zero mode, either by giving the system a mass larger than $\mu$, or by manually making $E_0$ real, the only instability in the system would be from the infinite energy states, and the system would actually heat to infinity for some initial conditions. In the presence of the imaginary zero mode, the decay always dominates at long timescales. However, since $\Delta E_m$ approaches zero from below as $m$ increases, it is possible for $U$ to have finite-energy eigenstates, as long as their weight is not on the lowest few energy states. Because of this, there are many stable states in the system at large momenta.
 Using the truncated matrix $U_{i,j}$ with $i,j\leq 4000,$ we can approximate the eigenvectors of $U$ through numerical diagonalisation. Although the propagator is not unitary, due to the decaying zero mode, the procedure still works because the propagator becomes diagonal at large energies. From the numerics we can distinguish two kinds of eigenvectors, those with unit norm eigenvalue, and those with an eigenvalue norm deviating from unity. A typical example of an eigenvector with unit norm is depicted in Fig.~\ref{UNev}.
In Fig.~\ref{UNev}(a), the $1024^{th}$ eigenvector, having its highest weight on $|f_{1024}^0\rangle$, is shown. It has a strong weight on a single eigenstate of $H^0$, with a quick decay away from the peak. A resonant structure is visible in the weights of the eigenvector, with peaks appearing at a spacing of $20$. The decay of the peaks is rapid, so a vertical filling has been used in the graph as a guide to the eye. In Fig.~\ref{UNev}(b), where a zoom of Fig.~\ref{UNev}(a) is shown, the peaks can be seen to consist of more than one point, although the decay is very sharp. The spacing between the peaks is precisely the driving frequency, so the resonant behaviour of the driving is also visible in the eigenvectors. The sharpness of the resonant peaks shows that the system efficiently excites the resonant states, while coupling the off-resonant states only very weakly. Finally, the tails of the eigenstates decay faster than $1/n$ so that the mean energy in such an eigenstate is finite. Numerical results indicate that it is roughly equal to the energy at the peak weight, which is consistent with the slower decay of the resonances to the left.

\begin{figure}[hbt]
\includegraphics[width=\linewidth]{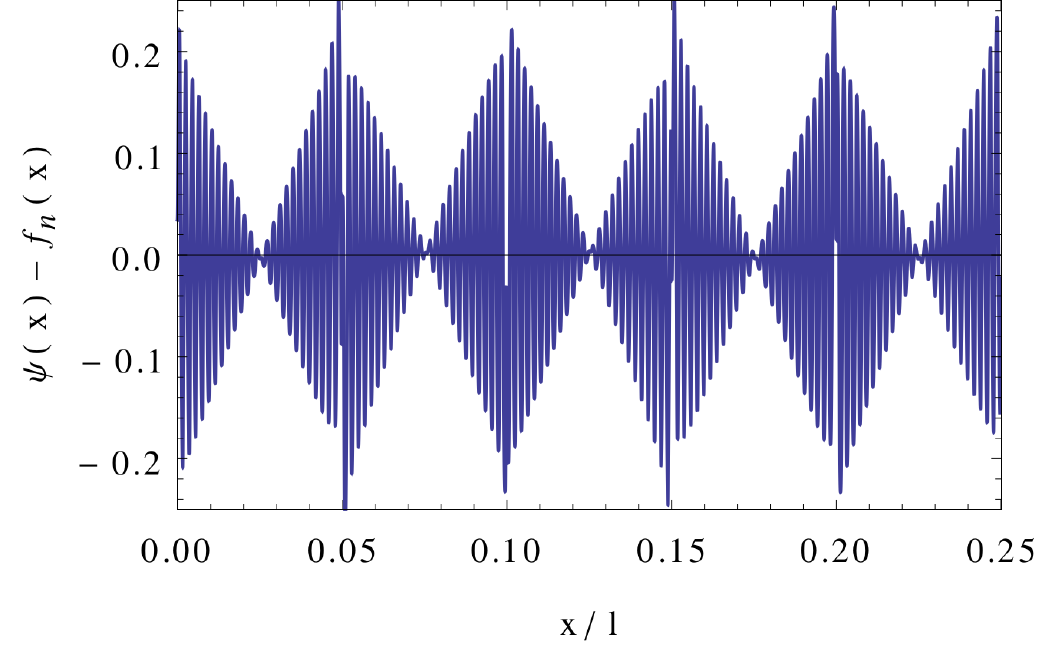}
\caption{\label{Beats} The real part of $\psi(x)-f_{1024}(x)$ is shown for the first quarter of the interval. The function $f_{1024}(x)$ is substracted to display the various beat frequencies present in $\psi(x)$. The beat has a period of $1/20$, which is the driving frequency, as expected.}
\end{figure}

Although the energy in the state remains roughly the same before and after driving, the resonances in the quasi-stable states have a measurable effect on the wavefunctions. Let $v$ be the $1024^{th}$ eigenvector, whose weights are shown in Fig.~\ref{UNev}. This state has the majority of its weight on the eigenfunction $f_{1024}(x)$ of the undriven Hamiltonian $H_0.$ Let $\psi(x)$ denote the wavefunction of $v$ in position space. To show the effect of the resonances, the real part of the function $\psi(x)-f_{1024}(x)$ is shown in Fig.~\ref{Beats}. By substracting the wavefunction $f_{1024}$ from the parent state, the presence of beat frequencies in the real part of $\psi(x)$ becomes visible. The beats repeat at a multiple of the driving frequency, as is expected from using the addition law for trigonometric functions. We do not show the imaginary part because it does not beat with the parent state (which has real wavefunction), and merely oscillates at very high frequency.

\begin{figure}[hbt]
\includegraphics[width=\linewidth]{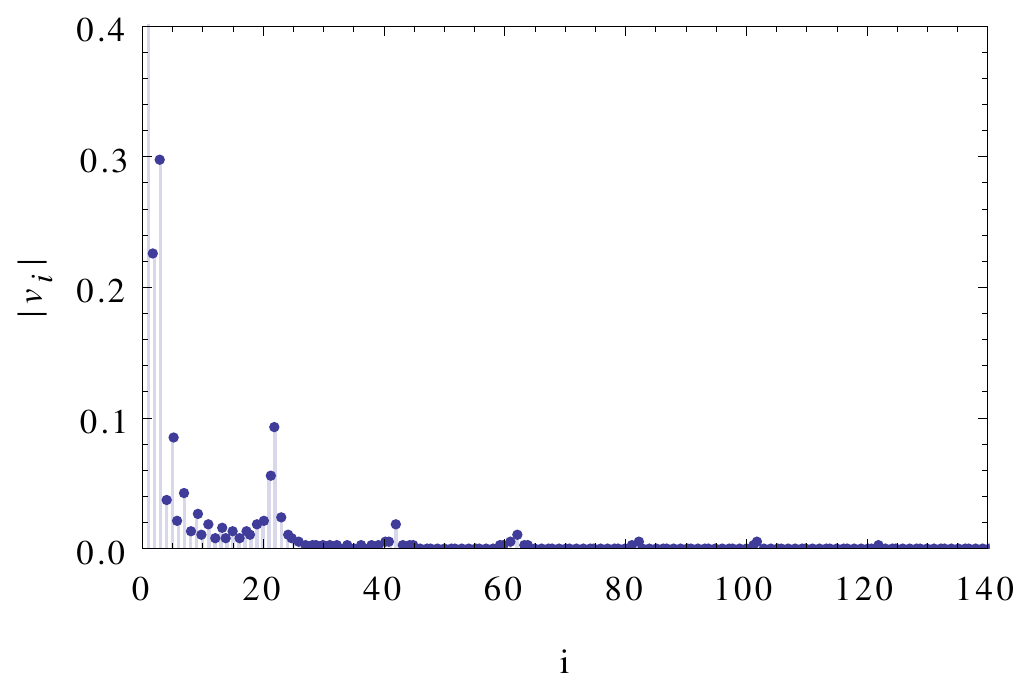}
\caption{\label{NUNev} The components $|v_i|$ of an eigenvector $v$ different from that in Fig.~\ref{UNev}, but for the same parameters. The corresponding eigenvalue has norm smaller than unity. The peak weight is one of the eigenstates that gains energy over a cycle.}
\end{figure}
There are also the eigenvectors corresponding to an eigenvalue that is not of unit norm, of which an example is depicted in Fig.~\ref{NUNev}. These eigenvectors have most of their weight on the low-lying states, and their eigenvalues all have norm smaller than one. The eigenvalues of these eigenvectors cannot have unit norm because there is energy emitted out of these states. This happens because states slowly lose their momentum due to the driving protocol, and when they start to significantly overlap with the bound state $f_0$, they will decay out of the system. Because the propagator becomes diagonal at large momenta, the high-momentum eigenstates do have unit norm, as is consistent with the results from Fig.~\ref{dE}. The system is prevented form gaining arbitrary amounts of energy because the propagator becomes diagonal at large momenta, and because there is the decaying zero mode. This prevents arbitrary heating of the system. Therefore, all the singular behaviour due to the driving occurs at low energies.

\section{Stability analysis}\label{stability}

We now discuss the stability of our results under the various choices that we have made to generate the results described above.

First, we shall consider the impact of regularising the Hamiltonian with a small mass, like we did in sec.~\ref{finite}. What ensures the stability of the Hamiltonian is that the overlaps between the eigenfunctions at different $\mu$ become diagonal at large energies, together with the presence of the decaying zero mode. For this effect, the precise value of the lowest eigenvalues is irrelevant. The small mass therefore has no effect here. However, as we mentioned before, $U$ also has eigenvectors for which $\langle v|H|v\rangle=\infty$. This implies that we can write
\begin{align*}
	|v\rangle=\sum_n v_n |f^0_n\rangle,
\end{align*}
where $v_n$ decay faster than $n^{-1}$ but slower than $n^{-3/2}$. Because these states have such long tails, they lead to errors in the numerical procedure. We obtain our numerical results by considering the truncated matrix $U_{i,j}$ with $i,j\leq 4000,$ and this neglects the tails of these states. The mass breaks the periodicity in the eigenvalues of the Hamiltonian at low momenta, and causes the coefficients $v_n$ to decay slightly faster. This causes the numerical determination of the eigenstates of $U$ to be more accurate, which allowed us to use a cutoff of $4000$, rather than even higher. The reason for the introduction of $M$ is therefore purely technical. Nevertheless, making the mass large enough has an additional interesting effect. Initially, it reduces the decay rate of $|f_0\rangle$, and reduces the rate at which it is populated, lowering the rate at which probability leaks out of the system. However, when $M>\mu$, $|f_0\rangle$ is actually stable, altering the qualitative behaviour of the system.  Since we have chosen $M$ much smaller than $\mu$, these effects play a negligible role, and our results should accurately describe the system at $M=0$.

Secondly, we consider increasing the driving strength $\mu$, which will cause the overlap of eigenfunctions to become diagonal less quickly, increasing the size of $U$ that must be used in the numerics, but not significantly changing the results. A similar consideration applies when one drives at a larger resonant frequency. Modifying the driving frequency away from resonance has the opposite result. The results above are for a driving frequency resonant at $20$ energy spacings. If one makes the driving frequency off resonant, the resonant peaks shown in Fig.~\ref{UNev} become less pronounced, improving convergence.

Finally, it is not necessary to choose $T_1=T_2$ in Eq.~\eqref{DrivingHam}. Choosing these two times to be unequal will result in an asymmetry in Fig.~\ref{UNev}, with the resonance peaks on one side becoming larger than on the other. Otherwise, no significant change in the results is visible. On the whole, this shows that the general features we describe in Sec.~\ref{finite} are robust, and do not depend on the specific choices we have made.

Now, the question of stability under long-term driving remains. In general, when one starts driving the system according to the protocol in Eq.~\eqref{DrivingHam}, the system will not be in an eigenstate of the Floquet propagator. As long as the initial state has no overlap with the low-momentum states of the system, the system is relatively stable. It will not heat, and it will only slowly lose a small part of its energy through the decaying state, because the coupling to low-momentum states is negligible. Physically, one can prepare a stable state quite easily. The high-momentum states in the system have negligible overlap with the infinite-energy states, except at the resonant peaks. By preparing the system in a high-momentum state that is off-resonant with the decaying state, one gets a stable time-evolution. In this case, the initial state will consist mostly of the Floquet eigenstate peaking at the same momentum value, and the driving will create beat frequencies in the wavefunctions, while leaving the energy behaviour mostly untouched.

This situation becomes different when an uncertainty in the driving protocol is considered. In the above, we analysed the stability of our results under changes in the driving parameters; the presence of a random noise each Floquet cycle will have different results. It might be tempting to draw an analogy with spatial disorder, and hypothesise the onset of a kind of Anderson localisation in energy space. However, evidence indicates that this is not the case \footnote{A detailed analysis to this effect will appear in a future publication.}. Rather, what one can expect is a diffusion in energy space. This can be intuitively seen as follows: the propagator over a disordered driving can be written as $\delta U U_0$, where $U_0$ is a fixed unitary operator and $\delta U$ is a random unitary matrix. The random $\delta U$ will spread out a wavefunction in energy space, and the amount of spreading will be determined by the disorder. Iterating this process will create a system reminiscent of a random walk, and diffusion will result. This means that under disorder, the system would heat in the long run if the decaying mode were not present; this occurs because it has a lowest energy state, but no highest one. The diffusion downward is therefore limited, while the diffusion upward will continue arbitrarily. A recent result for the Ising model with disordered driving is consistent with this analysis \cite{Bhattacharya2017}. In the presence of the decaying mode, this diffusive behaviour would result in a continual population of the decaying mode, which would cause the particle to decay out of the system.

\section{Conclusion}\label{conclusion}

In conclusion, we have constructed a driven system that demonstrates various regimes. At high momenta, the system is stable under the driving; no energy is pumped into or out of the system due to the driving protocol, and the wavefunctions develop beat frequencies due to resonances in the driving. In contrast, the states at low momenta are unstable under the driving protocol, they lose their energy due to decay out of the system. However, because the driving over a period is local in momentum space, these two regimes can be separated. By initialising the system with only high-momentum states, there is no overlap with the decaying state, and the system should be stable under the driving. This shows that a careful choice of the driving protocol, which incorporates stimulated emission and stimulated absorptions in equal measures can lead to a quantum system that is actually stable under a periodic driving, even in the absence of integrability. The notion that heating rates can be controlled to some extent by choosing an efficient driving protocol is relevant in the context of optical lattices, where the experimental realisability of a variety of interesting phases is limited by heating effects.

A possible future area of research would be the inclusion of dissipation in the system. This would cause the system to decay out of the high-momenta states, while stabilising the ones at low momentum. This might cause stable states to develop at the transition between these two regimes. Another interesting possibility would be the inclusion of disorder, and interactions. These will interfere with the transmission of energy through the system, and will likely influence the heating properties in interesting ways.

We would like to thank Jean-S\'ebastien Caux for the many useful discussions that have contributed to this project. The work by A.Q. and C.M.S. is part of the D-ITP consortium, a program of the Netherlands Organisation for Scientific Research (NWO) that is funded by the Dutch Ministry of Education, Culture and Science (OCW).

\bibliographystyle{apsrev4-1}

\onecolumngrid
\appendix

\section{Asymptotic form of the propagator}\label{As}

Here we prove the asymptotic form 
\begin{align}
	U^{o,e}_{n+a,n}=2\theta e^{i\pi \theta(4n+a)}\operatorname{sinc}(a\pi\theta),
\end{align}
for the even and odd parts of the propagator, $U_e$ and $U_o$, respectively.

By performing the sum in Eq.~\eqref{Propagator}, the matrix elements for the odd part of the propagator can be written
\begin{align*}
	U^o_{m,n}=&\frac{1}{\pi^2(n-m)(n+m-1)}e^{i\pi\theta(2n+1)}(2m-1)\left(\Phi_z(1,\frac{1}{2}+m)-\Phi_z(1,\frac{3}{2}-m)\right)\\
	+&\frac{1}{\pi^2(n-m)(n+m-1)}e^{i\pi\theta(2n+1)}(2n-1)\left(\Phi_z(1,\frac{3}{2}-n)-\Phi_z(1,\frac{1}{2}+n)\right),
\end{align*}
where $m\neq n$, $z:=\exp(2i\pi \theta),$ and $\Phi$ is the Lerch transcendent. If $m=n$ we get
\begin{align*}
	U^o_{n,n}=\frac{1}{\pi^2(2n-1)}e^{i\pi\theta(2n+1)}(2m-1)&\left[2\Phi_z(1,\frac{3}{2}-n)-2\Phi_z(1,\frac{1}{2}+n)\right.\\
	+&\left.(2n-1)\left(\Phi_z(2,\frac{3}{2}-n)-\Phi_z(2,\frac{1}{2}+n)\right)\right].
\end{align*}
Using the recursion relation $z\Phi_z(s,a+1)=\Phi_z(s,a)-1/a^s$, we find that the Lerch transcendent becomes independent of $a$ at large $a$, up to a phase. Through numerical evaluation we find $|\Phi_z(s,\inf)|=0$, $|\Phi_z(s,-\inf)|=c_s$, where $c_s$ is independent of $z$, but not of $s$. Including the phase $z$ from the recursion relation, we find $U^o_{m+1,n+1}=z^2 U^o_{m,n}$ for $m,n$ large. This periodicity is a residue of the conformal symmetry. Far from the ground state, the system cannot distinguish between individual energy levels due to the absence of an absolute energy scale in the system. Using $c_1=\pi$, we can further simplify, to obtain for large $n$
\begin{align*}
	U^o_{n+a,n}=2\theta e^{i\pi \theta(4n+a)}\operatorname{sinc}(a\pi\theta)
\end{align*}
for $a\neq 0$, up to a global phase independent of $a,n,\theta$. The case $a=0$ can be evaluated in terms of $c_2$ explicitly if one wishes. 

We perform the same analysis for $U^e,$ and obtain
\begin{align*}
	U^E_{m,n}=\frac{-2^{-\delta_{m,0}/2}2^{-\delta_{n,0}/2}}{2\pi^2mn(n^2-m^2)}&\left(e^{2i\theta\pi(m+n)}n\left[B_z(\frac{1}{2}-m,0)+8B_z(\frac{3}{2}-m,-2)\right]\right.\\
	&-\left.e^{4i\theta\pi n}m\left[B_z(\frac{1}{2}-n,0)+8B_z(\frac{3}{2}-n,-2)\right]\right.\\
	&+\left.m\left[B_z(\frac{1}{2}+n,0)+8B_z(\frac{3}{2}+n,-2)\right]\right.\\
	&-\left.e^{2i\theta\pi (n-m)}n\left[B_z(\frac{1}{2}+m,0)+8B_z(\frac{3}{2}+m,-2)\right]\right),
\end{align*}
where $B_z(a,b)$ is the Incomplete Beta function. Using the periodicity property 
\begin{align*}
B_z(a,b)=\frac{1}{a+b-1}\left[(a-1)B_z(a-,b)-(1-z)^bz^{a-1} \right], 
\end{align*}
together with the asymptotic behaviour, ${B_z(\inf,-2)/m^2=0}$, ${B_z(-\inf,-2)/m^2=c}$, we can derive the same periodicity and asymptotic as for $U^o$.

\section{Expression for the Energy flux}\label{T}

Here, we derive the expression for the momentum flux $T_{t,x}(0)-T_{t,x}(l)$ from the main text. We start from the expression $T_{t,x}(x)=\normOrd{\pi(\bm x)\phi_x(\bm x)}$ given in the main text. By using the expansion $\phi(\bm x)=\sum_n \phi_n(t) f_n(x)$ and the expression for $a_n$ given in the main text, we obtain the expressions
\begin{align*}
	\phi(\bm x)&=\sum_n \sqrt{\frac{\hbar v}{2 l E_n}}f_n(x)(a_n(t)+a_n^\dagger(t))\\
	\pi(\bm x)&=i \sum_n \sqrt{\frac{l E_n \hbar}{2 v}}f_n(x)(a_n^\dagger(t)-a_n(t)).
\end{align*}
This allows us to write
\begin{align*}
	\normOrd{\pi(\bm x)\phi_x(\bm x)}&=\frac{i\hbar}{2}\sum_{n,n'}f_{n'}(x)f'_{n}(x)\sqrt{\frac{E_{n'}}{E_n}}\normOrd{[a^\dagger_{n'}(t)-a_{n'}(t)][a_n(t)+a_n^\dagger(t)]}\\
	&= \frac{i\hbar}{2}\sum_{n,n'}f_{n'}(x)f'_{n}(x)\sqrt{\frac{n'}{n}}\left[a^\dagger_{n'}(t) a_n(t)-a_n^\dagger(t) a_{n'}(t)+a_{n'}^\dagger(t) a_n^\dagger(t)-a_{n'}(t)a_n(t)\right],
\end{align*}
which can be rearranged to give
\begin{align}\label{SE}
&T_{t,x}(\bm x)=\nonumber\\
&-\frac{i\hbar}{2}\sum_{n,n'}\sqrt{\frac{n'}{n}}\left\{\left[f_n(x) f'_{n'}(x)-f'_n(x)f_{n'}(x)\right]a^\dagger_n(t) a_{n'}(t)+f_n(x) f'_{n'}(x)a^\dagger_n(t) a^\dagger_{n'}(t) -f_n(x) f'_{n'}(x)a_n(t) a_{n'}(t) \right\}.
\end{align}
By using the Robin boundary conditions $f'(x)=\mu f(x)/l$ at the boundary, we see that the term proportional to $a^\dagger a$ vanishes, and the total flux into the system is
\begin{align*}
T_{t,x}(0,t)-T_{t,x}(l,t)=\frac{i\hbar\mu}{2 l}\sum_{n,n'}\sqrt{\frac{n'}{n}}\left[f_n(l) f_{n'}(l)-f_n(0) f_{n'}(0)\right]\left(a^\dagger_n(t) a^\dagger_{n'}(t) -a_n(t) a_{n'}(t)\right).
\end{align*}
This is the expression quoted in the main text, with the constant of proportionality included.

\end{document}